\documentclass[12pt,a4paper]{article}
\usepackage[dvips,final]{graphicx} 
\usepackage{tabularx} 
\usepackage{amsfonts}
\usepackage{amssymb,amsmath}

\def\du{\unskip\smash{\lower 1.4ex \hbox{\char34}}\kern-.2ex}
\def\hu{\kern-.2ex\hbox{\char92}}

\def\XXint#1#2#3{{\setbox0=\hbox{$#1{#2#3}{\int}$}
     \vcenter{\hbox{$#2#3$}}\kern-.5\wd0}}

\newcommand{\bdis}{\begin{displaymath}}
\newcommand{\edis}{\end{displaymath}}
\newcommand{\be}{\begin{equation}}
\newcommand{\ee}{\end{equation}}

\newcommand{\noi}{\noindent}

\newcommand{\HBM}{\hat{H}_M}

\begin{document}
\baselineskip=7.6mm

\title{False vacuum decay in a brane world cosmological model}
\author{Michal Demetrian\footnote{e-mail: demetrian@fmph.uniba.sk} \\
Comenius University, Mlynska Dolina M105 842 48 \\ Bratislava IV, Slovakia}

\maketitle

\abstract{
The false vacuum decay in a brane world model is studied in this work. We
investigate the vacuum decay via the Coleman-de Luccia instanton, derive explicit approximative
expressions for the Coleman-de Luccia instanton which is close to a Hawking-Moss instanton
and compare the results with those already obtained within Einstein's theory of relativity.
} \\
\noi


\section{Introduction}

The total energy-density of our universe is very close to the
critical value corresponding to the Friedmann-Robertson-Walker metrics of the flat type. Numerically
this means that the total matter-density parameter has the value in the thin range around $1$:
$\Omega=1.02\pm 0.04$. There are two possible interpretations of
this fact: the universe is exactly flat; or the early evolution forced the universe to evolve to the present state
which is very close to the flat Friedmann-Robertson-Walker metric, however the geometry may be both open or closed. There are
models within open inflationary universe scenario, e.g.\cite{litoy}, \cite{limita}, \cite{yama}
that can successfully lead to
an acceptable present-day universe. These models are built on Einstein's theory of relativity in four
dimensional space-time. There are also models of "creation of an infinite universe within a finite bubble"\ based
on modifications of the Einstein's relativity theory like the model presented in
\cite{cahe1} where the false vacuum decay via
Coleman-de Luccia (CdL) instanton and subsequent second phase of inflation within the nucleated bubble is studied
in the context of Jordan-Brans-Dicke theory. \\
The progress in the superstring theory during last years has forced the cosmologists to consider
the extra dimensions in various models of the universe evolution e.g. \cite{gata} and \cite{gasa}. Especially, the
gravity-scalar instantons have been considered in \cite{mbl}.
The creation of
an open or closed universe within the brane world scenario has been considered in \cite{bouchmadi}.
The physical and geometrical discussion of the semiclassical instability of the Randall-Sundrum brane world resulting in
the vacuum decay via instantons is done in \cite{ochiai}, \cite{padilla}. \\
In this paper we consider the decay of the false vacuum of a
scalar field (inflaton) confined to a four dimensional brane in a
five dimensional brane world model. We analyze the Euclidean
cosmological equations in the Randall-Sundrum type II scenario
\cite{randall} that are supposed to describe semiclassically the
false vacuum decay. We are inspired by the work done by del Campo,
Herrera and Saavedra \cite{cahesa} in which the authors
investigate the possibility of realization of the open inflation
scenario in the brane world models including the existence of the
Coleman-de Luccia instanton providing the false vacuum decay and
subsequent inflation within created open universe. The authors of
the cited paper are interested in a specially chosen theory (the
self interaction $V$ of the scalar field). They investigate the
model only for fixed parameters of the self-interaction and obtain
both CdL instanton and plausible evolution after tunneling. Our
aim is to study the CdL instantons for arbitrary potentials and
compare the properties of CdL instantons in the standard
Einstein's gravity in four dimensional space-time with those from
our brane world. A similar problem has been recently considered in
\cite{brechet}, where the authors study the vacuum decay on the
brane within the thin-wall approximation. Unlike the authors of
the paper \cite{brechet} we are interested in another problem that
can be solved analytically, namely the CdL instanton(s) of the
first order close to a Hawking-Moss instanton. \\
The paper is organized as follows: in section 2 we briefly review the basic fact about the CdL instantons
in four dimensional de Sitter space-time to be able to
compare them with the results of this paper. In section three we present the formulation of the instanton equations in
considered brane world model and some consequences of these equations are discussed. Perturbative computation of the
first order CdL instanton in our brane world model is presented in section four, and finally the main quantity
characterizing the instanton - its action - is computed in the fifth section.

\section{False vacuum decay via CdL instanton in four dimensional space-time within Einstein's relativity theory}

CdL instanton introduced in \cite{cdl} describes false vacuum decay in a de Sitter universe within the
semiclassical approximation.
If $V=V(\Phi)$ is the effective potential for the scalar field,
this CdL instanton can be introduced as the $O(4)$-symmetric and finite-action solution of the Euclidean version of the
Einstein equations. The $O(4)$-symmetry means that the scalar field $\Phi$ lives on a (squeezed) four-sphere with
the metric
\be \label{euclmet}
{\rm d}s^2={\rm d}\tau^2+a^2(\tau)\left[{\rm d}\chi^2+\sin^2(\chi){\rm d}\Omega_2^2\right]  ,
\ee
with $\tau\in[ 0,\tau_f]$. The action (Euclidean version of the Einstein-Hilbert action) reads
\begin{eqnarray} \label{eucact}
& &
S= \nonumber \\
& &
2\pi^2\int_0^{\tau_f}\left[\left(\frac{1}{2}(\Phi')^2+V\right)a^2-\frac{1}{C}(a(a')^2+1)\right]a{\rm d}\tau  ,
\end{eqnarray}
where $C=8\pi/3$ and the prime denotes the derivative with respect to $\tau$.
Varying the action (\ref{eucact}) we obtain the (Euclidean) equations of motion for the scale parameter and the
inflaton:
\be \label{ee1}
a''(\tau)=-C\left( (\Phi')^2+V\right)a,\ \Phi''+3\frac{a'}{a}\Phi'-\partial_{\Phi}V=0  .
\ee
The local energy conservation law and the requirement of finiteness of the action impose the boundary conditions on the
functions $a$ and $\Phi$
\be \label{bc}
a(0)=0,a'(0)=1, \Phi'(0)=\Phi'(\tau_f)=0  ,
\ee
where $\tau_f>0$ is to be determined from $a(\tau_f)=0$. The action (\ref{eucact}) of a CdL instanton can be
considerably simplified by using the equations of motion (\ref{ee1}):
\begin{displaymath}
S=-\frac{4\pi^2}{3C}\int_0^{\tau_f} a{\rm d}\tau \ .
\end{displaymath}
After Coleman and de Luccia have proposed the idea of
the description of the vacuum decay in \cite{cdl}, many authors have studied the system (\ref{ee1}) and (\ref{bc}).
Coleman and de Luccia themselves have found the solution of the instanton equations in the thin-wall limit. We suppose
the effective potential $V$ is non-negative function with two non-degenerate minima.
The top of the potential barrier is reached at the point we denote by $\Phi_M$.
Furthermore, we denote: $V_M\equiv V(\Phi_M)$, $V_M''\equiv \partial^2_{\Phi}V(\Phi_M)$, $H_M^2=CV_M$ and
$H^2(\Phi)=CV(\Phi)$. The
properties of $V$ in the neighborhood of $\Phi_M$ are crucial for the existence of CdL instanton.
Motivated by the earlier works \cite{js} and \cite{tanaka} the authors of the papers \cite{jv1}, \cite{jv2} and
\cite{j3}
have achieved information about the solution(s) of the instanton equations interesting for us:
\begin{itemize}
\item if $-V_M''/H_M^2>4$ then the CdL instanton exists
\item if a CdL instanton exist then $-V''(\Phi)/H^2(\Phi)>4$ somewhere in the barrier
\item if $-V_M''/H_M^2>l(l+3)$, where $l$ is an arbitrary integer, then CdL instaton crossing in the $\Phi$-direction
$l$-times the top of the barrier  ($l$th order CdL instnton) exists
\item if $-V_M''/H_M^2\to l(l+3)$ we have the explicit approximative formulas
for the instanton and its action, \cite{jv2}
\end{itemize}
%
%
\section{False vacuum decay on a brane - elementary discussion of the instanton equations}

The dynamics of the scalar field confined to a four dimensional brane in a five dimensional bulk in our model is
defined by the action
\be \label{actionbrane}
S=M_{(5)}^3\int{\rm d}^5x\sqrt{|G|}\left[ R_{(5)}-2\Lambda_{(5)}\right]-\int{\rm d}^4x\sqrt{|g|}\mathcal{L}_\Phi
\ee
with the matter term
\bdis
\mathcal{L}_\Phi=\frac{1}{2}\partial_\mu\Phi\partial^\mu\Phi-V(\Phi) \ ,
\edis
where $R_{(5)}$ is the scalar curvature of the five-dimensional bulk metric $G$, $M_{(5)}$ and $\Lambda_{(5)}$
stand for the five-dimensional Planck mass and cosmological constant, respectively. These quantities are related
to the effective four-dimensional cosmological constant $\Lambda_{(4)}$, the brane tension $\sigma$ and
the four-dimensional Planck mass $M_{(4)}$ by the relations
\begin{align}  \label{reducteqs}
&\Lambda_{(4)}=\frac{4\pi}{M_{(5)}^3}\left( \Lambda_{(5)}+\frac{4\pi}{3M_{(5)}^3}\sigma^2\right),& \nonumber \\
&M_{(4)}=\left(\frac{3}{4\pi}\right)^{1/2}\frac{M_{(5)}^3}{\sigma^{1/2}}.&
\end{align}
We will use the units where $M_{(4)}=1$.
Following the works \cite{shiromizu} and \cite{cahesa} we come at the Euclidean equations of motion for the inflaton
$\Phi$ on the brane and the induced metric (under the assumption of $O(4)$-symmetry which involves the line
element of the form ${\rm d}s^2={\rm d}\tau^2+a^2(\tau)[{\rm d}\chi^2+\sin^2(\chi){\rm d}\Omega_2^2]$):
\begin{align} \label{eebrane}
&a''=-C\left\{ (\Phi')^2+V+\frac{1}{8\sigma}\left[(5(\Phi')^2+2V)(-(\Phi')^2+2V)\right]\right\}a,&
\nonumber \\
&\Phi''+3\frac{a'}{a}\Phi'-\partial_\Phi V=0.&
\end{align}
The functions
$a$ and $\Phi$ must obey the boundary conditions (\ref{bc}). We see that in the $\sigma\to +\infty$ limit we
recover the standard general-relativistic equation for the scale parameter $a$ \cite{cdl}, the equation for $\Phi''$
remains unchanged with respect to the Einstein's relativity theory.
If we assume that $(\tilde{a},\tilde{\Phi})$ is a CdL instanton with
$\Phi(0)=\Phi_i$ and $\tilde{\Phi}(\tau_f)=\Phi_f$, then we can write for $\tau\to 0^+$:
\be \label{epi}
\tilde{\Phi}''+\frac{3}{\tau}\tilde{\Phi}'-\partial_\Phi V(\Phi_i)=0 \Rightarrow
\tilde{\Phi}=\Phi_i+\frac{\partial_\Phi V(\Phi_i)}{8}\tau^2 ,
\ee
and for $\tau\to \tau_f^-$:
\be \label{epf}
\tilde{\Phi}''+\frac{3}{\tau}\tilde{\Phi}'-\partial_\Phi V(\Phi_f)=0 \Rightarrow
\tilde{\Phi}=\Phi_f+\frac{\partial_\Phi V(\Phi_f)}{8}(\tau_f-\tau)^2 .
\ee
Under the assumption that $a$ is a concave function (surely guaranteed by the positivity of the term
proportional to $1/\sigma$ in the equation for $a''$) we deduce from eqs. (\ref{epi}) and (\ref{epf}) that the
CdL instanton (in $\Phi$-direction) must cross the value $\Phi_M$ once at least.
The asymptotic of the non-instanton solution (we can say - solutions with "randomly"\ chosen
initial value of $\Phi$) of the system of equation (\ref{eebrane}) can be found in the same way as in Einstein's
general relativity.
\\
We are interested in solutions which are
close to the so-called Hawking-Moss (HM) instanton \cite{hm} that describes the false vacuum tunneling as a process
in which the inflaton "jumps"\ (within a horizon-size domain) at the top ($\Phi_M$) of the potential $V$.
The HM instanton is the $O(5)$-symmetric (constant $\Phi$) solution of the system (\ref{eebrane}):
\be \label{hmb}
\Phi=\Phi_M,\ a= \HBM^{-1}\sin\left( \HBM\tau\right)\ ,
\ee
where $\HBM$ is a modification of the Hubble parameter $H_M$ introduced in the previous section. $\HBM$ is
determined inserting the proposed solution into the first equation of (\ref{eebrane}). One easily obtains
\bdis
\HBM^2=\frac{8\pi}{3}\left( V_M+\frac{V_M^2}{2\sigma}\right)=H_M^2\left( 1+\frac{V_M}{2\sigma}\right) \ .
\edis
We can study the CdL instantons close to this HM instanton in the following way. We insert the expression for $a$ from
(\ref{hmb}) into the equation for $\Phi''$, linearize the term $\partial_\Phi V$ and using new
variables: $x=\HBM\tau$ and $y=\Phi-\Phi_M$ we obtain
\bdis
\frac{{\rm d}^2y}{{\rm d}x^2}+3\cot(x)\frac{{\rm d}y}{{\rm d}x}-\frac{V_M''}{\HBM^2}y=0 \ ,
\edis
or transforming the independently variable $x$ to: $z=\cos(\HBM\tau)$ we get
the standard hypergeometric equation:
\bdis
(1-z^2)\frac{{\rm d}^2y}{{\rm d}z^2}-4z\frac{{\rm d}y}{{\rm d}z}-\frac{V_M''}{\HBM^2}y=0 \ .
\edis
The boundary conditions (\ref{bc}) restrict possible values of the parameter $\frac{V_M''}{\HBM^2}$ to the
eigenvalues of the Laplace-Beltrami operator on $S^4$:
\be \label{propval}
-\frac{V_M''}{\HBM^2}=l(l+3),\ l\in\{ 0,1,2,\dots \}
\ee
and the solutions $y=y_l$ read for odd $l$:
\be \label{y1odd}
y_l=c_lz\ _2F_1\left(\frac{1-l}{2},2+\frac{l}{2},\frac{3}{2},z^2\right)
\ee
and for even $l$:
\be \label{y1even}
y_l=c_l\ _2F_1\left(\frac{3+l}{2},-\frac{l}{2},\frac{1}{2},z^2\right) \ ,
\ee
where $c_l$ are arbitrary constants and $_2F_1$ stands for nondegenerate hypergeometric function.
(In fact, the hypergeometric functions with special arguments according (\ref{y1odd}),(\ref{y1even}) reduce to
the Gegenbauer polynomials in the variable $z$. However, we will not need this explicitly.)
The function $y_0$ correspond to the HM instanton
and the functions $y_l$ approximate the $l$th order CdL instanton in its $\Phi$-direction.
The restriction (\ref{propval}) is formally the same as in the
case of four dimensional space-time with $H_M$ changed to $\HBM$. This change means that we have a new parameter
(except the old one $-V_M''/H_M^2$)
which value is crucial for the existence of the CdL instanton, namely $V_M/\sigma$. It is obvious that for
$V_M/\sigma\ll 1$ the theory of vacuum decay on our brane reduces to the theory of vacuum decay in four-dimensional
space-time.
The first-order CdL instanton plays the most important role in the Einstein's
theory of gravity \cite{tanaka}, \cite{jv2}, therefore
we write down explicitly:
\be \label{y1lin}
y_1=k z =k\cos\left( x\right)\ ,
\ee
with $k$ - the amplitude of the inflaton during its Euclidean evolution. In the next we will be interested in the
first-order CdL instanton only.

\section{The first order CdL instanton - perturbative approach}
The idea of our analysis of the system of equations (\ref{eebrane}) is to expand all the relevant quantities
entering these equations (and the boundary conditions (\ref{bc})) into the powers of the $\Phi$ amplitude $k$,
see (\ref{y1lin}). This means explicitly that the following formulas are of our interest:
\begin{eqnarray*}
& &
y(x)=\sum_{n=0}^\infty k^nu_n(x), \
a(x)=\HBM^{-1}\sum_{k=0}^\infty k^nv_n(x) \ , \\
& &
-\frac{V_M''}{\HBM^2}=4+\sum_{n=1}^\infty k^n\Delta_n ,
\end{eqnarray*}
together with the Taylor expansion of the potential $V$ around its local maximum.
It can be useful to write down the form of the system of linear equations by which we replace equations (\ref{eebrane}):
\begin{eqnarray*}
& &
u_n''+3\cot(x)u_n'+4u_n=\mathcal{U}_n,\ v_n''+v_n=\mathcal{V}_n\sin(x),
\end{eqnarray*}
where the source-terms $\mathcal{U}_n$ and $\mathcal{V}_n$ are to be computed order-by-order.
We know, from the previous section, that
\bdis
u_0=0,\ v_0(x)=\sin(x),\ u_1(x)=\cos(x) .
\edis
The boundary conditions imposed on $a$ and $\Phi$ require that for all $n\geq 1$ we have
$v(0)=v'(0)=0$. The right-end point $x_f$ at which the derivative of $y$ has to vanish is determined
by $a(x_f)=0$, therefore we have also the expansion of this quantity:
\bdis
x_f=\pi+\sum_{n=1}^\infty k^nx_f^{(n)} .
\edis
The fact that the potential $V$ is supposed to have the local maximum at $\Phi_M$ implies that
$v_1=0$ and $x_f^{(1)}=0$. The contribution of the second order in $k$ to the scale factor $a$ is nonzero, and reads
explicitly
\begin{eqnarray} \label{v2}
& &
v_2=\frac{1}{32}\left\{ \left[ C+\frac{15-9C}{4\sigma}V_M\right]4x\cos(x)+ \right. \nonumber \\
& &
\left.
\left[ 5C+\frac{51C-45}{4\sigma}V_M\right]\sin(x)-\right. \nonumber \\
& &
\left.
\left[ 3C+5(1+C)\frac{V_M}{4\sigma}\right]\sin(3x) \right\} .
\end{eqnarray}
Having this results we come at the shift of the right-end point $x_f$
\be \label{xf2}
x_f^{(2)}=-\frac{\pi}{8}\left[ C+\frac{3V_M}{4\sigma}(5-3C)\right] .
\ee
In the limit that $\sigma\to\infty$ this quantity is negative but for a finite value $V_M/\sigma$ it can be
both negative or positive and it vanishes at $V_M/\sigma\approx 0.555$.
Careful and a little bit tedious computation shows that $\Delta_1=0$ and that $u_2$ obeys equation
\bdis
u_2''+3\cot(x)u_2'+4u_2=\frac{1}{2}\frac{V_M'''}{\HBM^2}u_1^2 .
\edis
The right-end point is still $\pi$, i.e. we seek for the solution(s) for which $u_2'(0)=u_2'(\pi)=0$. This
determines $u_2$ as follows
\be \label{u2}
u_2=\frac{1}{24}\frac{V_M'''}{\HBM^2}\left[ 1-2\cos^2(x)\right] .
\ee
We continue with time-consuming computations without any extra-idea and derive the equation for $v_3$ we do not write
down. However,  the explicit formula for $v_3$ is
\begin{eqnarray} \label{v3}
& &
v_3=-\frac{V_M'''}{288\HBM^2}\left\{ 2C\left[ -2\sin(2x)+\sin(4x)\right]+\right. \nonumber \\
& &
\left.
\frac{V_M}{\sigma}\left[ -16(C-1)\sin(x)+2(3C-5)\sin(2x)+\right. \right. \nonumber \\
& &
\left. \left.
(C+1)\sin(4x)\right] \right\} .
\end{eqnarray}
We have divided the expression for $v_3$ into two parts: the first one does not contain the brane tension
$\sigma$ and represents the contribution coming from the Einstein's general relativity and the second one
is connected with the brane-tension containing terms of the action (\ref{actionbrane}). Finally, it remains to
find the function $u_3$ to determine
the amplitude of $\Phi$. First of all we should realize that the shift of the right-end point
(\ref{xf2}) as well as the term $k^2\Delta_2$ from the expansion of $-V_M''/\HBM^2$ enter the equation for $u_3$. This
fact allows for determining a relation between $k$ and $-V_M''/H_M^2$ as it is shown bellow. To keep the
range of the argument of function $u_3$ equal to $[0,\pi]$ we pass from the independent variable $x$ to
$w=( 1-k^2x_f^{(2)}/\pi)x$.
For the simplicity we introduce the notation
\begin{eqnarray*}
& &
E=C+\frac{15-9C}{4\sigma}V_M, \
F=5C+\frac{51C-45}{4\sigma}V_M, \\
& &
G=3C+5(1+C)\frac{V_M}{4\sigma}\ .
\end{eqnarray*}
Within this notation we can derive straightforwardly differential equation for $u_3$:
\be \label{u3eq}
u_3''+3\cot(w)u_3'+4u_3=A\cos(w)+B\cos^3(w) ,
\ee
where
\begin{eqnarray*}
& &
A=\frac{5}{8}C+\frac{3}{8}E+\frac{3}{4}G+\frac{1}{24}\left(\frac{V_M'''}{\HBM^2}\right)^2-\Delta_2 , \\
& &
B=-\frac{3}{4}G-\frac{1}{12}\left(\frac{V_M'''}{\HBM^2}\right)^2+\frac{1}{6}\frac{V_M''''}{\HBM^2}.
\end{eqnarray*}
The solution is: $u_3=\beta\cos^3(w)$,
where the coefficient $\beta$ has to satisfy two conditions:
\bdis
6\beta=A,\ -14\beta=B .
\edis
However, the fixation of $\beta$ is, at the moment, only supplementary for us because the coefficient $A$ contains
also $\Delta_2$.
Eliminating $\beta$ from the previous system of equations we obtain the expression for $\Delta_2$ (we do not
write down)
and subsequently we get $k^2$ as the function of $-4-V_M''/\HBM^2$:
\begin{eqnarray}  \label{k2brane}
& &
k^2=-7\left( 4+\frac{V_M''}{\HBM^2}\right)
\left\{ 16C+\frac{1}{24}\left(\frac{V_M'''}{\HBM^2}\right)^2+ \right. \nonumber \\
& &
\left.
\frac{1}{2}\frac{V_M''''}{\HBM^2}+\frac{1}{32}\frac{V_M}{\sigma}(435-69C)\right\}^{-1} .
\end{eqnarray}
Assuming $4+V_M''/\HBM^2<0$ we need positive sign of the denominator in eq. (\ref{k2brane}). This sign can be changed,
for given values of $V_M$ and $\sigma$, only due to $V_M''''/\HBM^2$.  We introduce the critical value of
$V_M''''/\HBM^2$
\be \label{zetac}
\zeta_c=
\frac{1}{16}\frac{V_M}{\sigma}(69C-435)-32C-\frac{1}{24}\left(\frac{V_M'''}{\HBM^2}\right)^2
\ee
at which the mentioned denominator vanishes, and we have the first-order CdL instanton with inflaton amplitude
given by the formula (\ref{k2brane}) in the two cases:
\begin{itemize}
\item
for $V_M''''/\HBM^2>\zeta_c$  as $4+V_M''/\HBM^2\to 0^-$ ,
\item
for $V_M''''/\HBM^2<\zeta_c$  as $4+V_M''/\HBM^2\to 0^+$ .
\end{itemize}
The first term in (\ref{zetac}) is positive (if $V_M$ and $\sigma$ are positive). This cause the difference with respect
to the situation when vacuum decays in four dimensional space-time within Einstein's general relativity because now
$V_M''''/\HBM^2$ can be both positive and less than $\zeta_c$
(In the first paper of refs. \cite{j3} it is argued that the negative value of $V_M''''$ is not like).
Let us mention that the situation when $-V_M''/\HBM^2>4$ and $V_M''''/\HBM^2<\zeta_c$
does not mean automatically
a kind of stabilization of the false vacuum because we still can have
some CdL instanton with large amplitude in $\Phi$ that is not included in our previous analysis and
moreover the false vacuum can decay also via HM instanton.
However,
a consideration of such an instanton would involve some class of non-perturbative analysis of the system (\ref{eebrane}) or
the numerical analysis, if we have a concrete potential $V$ and a brane tension $\sigma$.

\section{The action of the first-order CdL instanton}

The crucial quantity that tells us how probable is the vacuum decay via the CdL (or HM) instanton is its
action. Our task is to find the approximative formula for the action of the
first-order CdL instanton we investigated in previous section. In \cite{cahesa} the authors have also
considered the action of the CdL instanton. They were interested in the action within the thin-wall approximation
that correspond with their example of the instanton. Following \cite{cahesa} we can write the action
\begin{eqnarray*}
& &
S=2\pi^2\int_0^{\tau_f}{\rm d}\tau
\left[a^3\left(\frac{1}{2}(\Phi')^2+V\right)+\frac{a^3}{2\sigma}\left(\frac{1}{2}(\Phi')^2+V\right)^2+\right. \\
& &
\left. \frac{1}{C}\left( a^2a''+a(a')^2-a\right)\right] .
\end{eqnarray*}
Using the equation of motion (\ref{eebrane}) we rewrite the action into a much simpler form
\begin{eqnarray} \label{acbra}
& &
S=-\frac{4\pi^2}{3C}\int_0^{\tau_f}a{\rm d}\tau-
\frac{\pi^2}{\sigma}\int_0^{\tau_f}a^3(\Phi')^4{\rm d}\tau= \\
& &
-\frac{4\pi^2}{3C}S^{(I)}-\frac{\pi^2}{\sigma}S^{(II)} .
\end{eqnarray}
The structure of the $S^{(I)}$ term is the same as in the case of Einstein's relativity theory. To avoid a confusion
we must mention that brane-tension is included in this term throughout $a$.
We can easily find the action of the HM instanton
\be \label{hmaction}
S_{HM}=-\frac{\pi}{\HBM^2}=-\frac{3}{8V_M}\frac{1}{1+\frac{V_M}{2\sigma}} .
\ee
We see that the action of the
HM instanton in our brane world model is for fixed $V_M$ and any given (positive) $\sigma$ less than the
action of corresponding HM instanton in Einstein's relativity theory.
Now we can compute the action of the first-order CdL instanton. Up to the
second order in $k$ one has
\begin{eqnarray*}
& &
S^{(I)}=\HBM^{-2}\left\{ \int_0^{\pi+k^2x_f^{(2)}} v_0{\rm d}x+k^2\int_0^\pi v_2{\rm d}x\right\}= \\
& &
\HBM^{-2}\left[ 2+k^2\frac{4C-5}{3}\frac{V_M}{\sigma}\right]
\end{eqnarray*}
and $S^{(II)}$ does not contribute because it is of the order $k^4$ at most.
Putting these results together we obtain the difference between the actions of our first-order CdL instanton and
related HM instanton
\begin{eqnarray} \label{actdiff}
& &
S_{CdL}-S_{HM}=-\frac{4\pi^2}{3C}\frac{4C-5}{3}\frac{1}{\HBM^2}\frac{V_M}{\sigma}k^2 .
\end{eqnarray}
First of all: this difference vanishes as $\sigma\to\infty$ as it must be because in Einstein's relativity theory
the difference between the actions of our CdL instanton and related HM instanton is of the fourth-order in $k$ as
it is shown in \cite{jv2}. For positive $k^2$ the difference (\ref{actdiff}) is negative (this coincides with
the general-relativistic result \cite{jv2}), this means that the vacuum decay via our CdL instanton is more
probable than the decay via HM instanton. The fact that the difference of the actions
(\ref{actdiff}) is of the $k^2$ order unlike the general-relativistic case where it is of the order $k^4$ means
that the false vacuum decay rate in our brane world model is higher than for the conventional gravity. The same is
shown for the false vacuum decay rate in another extremal case when the thin-wall approximation can be used
\cite{brechet}.

\section{Summary}

We have compared, from the point of view of the semiclassical description of the false vacuum decay,
the systems of differential equations (\ref{ee1}) and (\ref{eebrane}) defining the CdL instantons in
the standard Einstein's relativity theory and in the brane world model. We have been interested in the situation when
the effective curvature of the potential at its top is close to the critical value $4$. In this situation we were able
to obtain explicit perturbative formulas for the instanton, and mainly for its actions that is the relevant
quantity describing the instanton as "mediator" of the vacuum decay. This allows for comparing our CdL instanton with
always existing Hawking - Moss instanton. We have concluded that it is the first order CdL instanton that is
preferred (to the corresponding Hawking - Moss instanton) in the considered region of parameters of the potential.
A kind of lower symmetry of the instanton equations with respect to the situation in general relativity causes that the
difference between actions of mentioned instantons is proportional to the square of inflaton amplitude in the instanton
rather than to the fourth power of that amplitude.  Physically, this difference of the actions is
inversely proportional to the string tension ($V_M/\sigma$ is the parameter controlling the string tension influence) and
therefore in the limit $V_M/\sigma\to 0$ we recover the results of the general relativity.

\subsubsection*{Acknowledgement}
This work was supported by the Slovak Scientific and Educational Grant Agency, project no. 1/0250/03.

\end{document}